\documentclass[smallextended]{svjour3}       
\smartqed  
\usepackage{graphicx}
 \journalname{Experimental Astronomy}
\begin{document}

\title{The Gaia Survey Contribution to EChO Target Selection and Characterization
}

\titlerunning{The Gaia Contribution to EChO}        

\author{Alessandro Sozzetti         
\and
        Mario Damasso
}

\authorrunning{A. Sozzetti \& M. Damasso} 

\institute{A. Sozzetti \at
              INAF - Osservatorio Astrofisico di Torino, \\
              Via Osservatorio 20, I-10025 Pino Torinese, Italy \\
              Tel.: +39-011-8101923\\
              Fax: +39-011-8101930\\
              \email{sozzetti@oato.inaf.it}           
           \and
           M. Damasso \at
              INAF - Osservatorio Astrofisico di Torino, \\
              Via Osservatorio 20, I-10025 Pino Torinese, Italy 
}

\date{Received: date / Accepted: date}

\maketitle

\begin{abstract}

The scientific output of the proposed EChO mission (in terms of spectroscopic 
characterization of the atmospheres of transiting extrasolar planets) will be 
maximized by a careful selection of targets and by a detailed characterization 
of the main physical parameters (such as masses and radii) of both the planets 
and their stellar hosts. To achieve this aim, the availability of high-quality data from other 
space-borne and ground-based programs will play a crucial role. Here we identify 
and discuss the elements of the Gaia catalogue that will be of utmost 
relevance for the selection and characterization of transiting planet 
systems to be observed by the proposed EChO mission.


\end{abstract}

\section{Introduction}\label{intro}

Gaia (e.g., Lindegren 2010), successfully launched on December 19, 2013, in its five-year all-sky astrometric survey 
will deliver direct parallax estimates for nearby main-sequence stars down to the broadband G=20 mag limit of the survey. 
At V=15 mag, typical F-G-K dwarfs within 0.5 kpc from the Sun will have parallaxes measured to better  than 1-2\% accuracy. 
Distances to relatively bright ($K < 10$) early- to late-M stars out to 20-30 pc will be known with better than 0.1\%-1\% 
precision (depending on spectral sub-type). This will constitute an improvement of up to over a factor 100 with respect 
to the typical 25\%-30\% uncertainties in the distance reported for low-mass stars identified as nearby based on proper-motion 
and color selections (e.g., L\'epine 2005; L\'epine \& Gaidos 2011). 

Furthermore, Gaia is equipped with two low-resolution prism spectrophotometers which together provide the spectral energy distribution 
(SED) of all targets in the range $330-1050$ nm, with a resolution varying between 13 and 85. The main purpose of these spectrophotometers 
(named BP and RP, for ``blue photometer'' and ``red photometer'') is to classify the objects (into star, galaxy, quasar etc.) and to 
odetermine the stars' physical properties (as well as to provide a chromaticity correction for the astrometric data analysis). 
For stars at $G=15$ (in Gaia's magnitude system) with less than two magnitudes of extinction, the expectation is that based solely on the BP/RP 
spectra it will be possible to estimate $T_\mathrm{eff}$ to within 1\%, $\log g$ to 0.1-0.2 dex, and [Fe/H] (for FGKM stars) to 0.1-0.2 dex (Liu et al. 2012). 

Finally, the single-field-of-view-transit photometric standard errors of the integrated $G$-band (Jordi et al. 2010) will remain 
below 1 mmag for $G < 13$. At $G=15$, they are expected to be on the order of 2 mmag, with only a weak color-dependence. The $G$-band 
data will be of particular use for stellar variability studies.

\section{Planet Hosts Characterization}\label{sec:1}

\begin{figure}
  \includegraphics[width=1.0\textwidth]{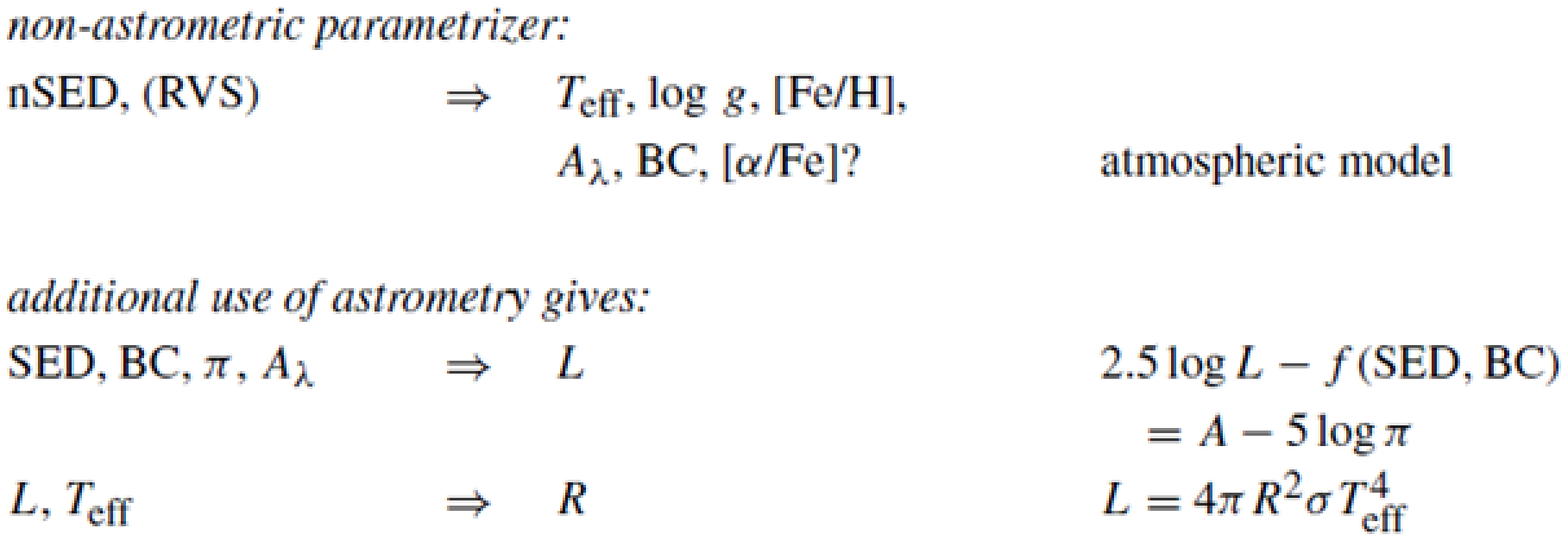}
\caption{Stellar parameters derivable from the Gaia data. SED=spectral energy distribution (spectrophotometric measurements in 
medium and broad band filters); nSED=normalized SED (absolute flux information removed); RVS=radial velocity spectrum; 
BC=bolometric correction; $\pi$=parallax; $A_\lambda$=interstellar extinction function. From Table 1 in Bailer-Jones (2002).}
\label{fig:1}       
\end{figure}
Starting with early data releases around mid-mission (astrometric parameters for most of Gaia stars will be made available as early 
as 28 months after mission launch, in mid-2016), the Gaia exquisitely precise distance estimates, and thus absolute luminosities, 
to nearby late-type stars will allow to improve significantly standard stellar evolution models at the bottom of the main sequence. 
For transiting planet systems, updated values of masses and radii of the host stars will be of critical importance. Model predictions 
for the radii of M dwarfs show today typical discrepancies of $\sim15\%$ with respect to observations, and as shown by the GJ 1214b 
example (Charbonneau et al. 2009) limits in the knowledge of the stellar properties significantly hamper the understanding of the 
relevant physical characteristics (density, thus internal structure and composition) of the detected planets. With Gaia, stellar 
radii for this sample are expected to be measured to within 2-3\% precision. However, the impact of Gaia parallaxes will be very 
relevant also for improved characterization of most of the F-,G- and K-type transiting planet hosts presently in the target list 
of EChO, the vast majority of them lacking a direct distance estimate (Note that final parallax estimates will be made available 
to the scientific community in the global Gaia catalogue 1 year prior to EChO's nominal launch date). 

In order to obtain precise radii estimates (a derived quantity), an approach similar to that described by Bailer-Jones (2002) might be adopted. 
In particular, the high-$SNR$ ($\sim200$) spectral energy distributions from the BP/RP spectra for bright ($V<15$), not heavily reddened stars will 
allow, in combination with the exquisitely precise parallaxes, to obtain accurate (to a few percent) absolute luminosities $L$. Provided the 
effective temperature $T_\mathrm{eff}$ estimates can be precise to 1\% (feasible for bright stars, see Liu et al. 2012), then, for the solar-type stellar 
sample targeted by EChO, the expectation is for a determination of stellar radii to better than 3\% precision. A summary of the process 
to derive stellar radii using Gaia data is given in Figure~\ref{fig:1}. In the eventuality that the $SNR$ of Gaia observations were to be reduced 
for a given star, thus not allowing to achieve the required precision on $T_\mathrm{eff}$ determinations, it is worth mentioning how all the presently 
confirmed transiting planets have been observed with high-resolution spectrographs at 4-10m telescopes, thus the objective can also be 
achieved by combining the absolute luminosity estimates from Gaia with effective temperature determinations using ground-based high-resolution spectroscopy. 

\begin{figure}
  \includegraphics[width=1.0\textwidth]{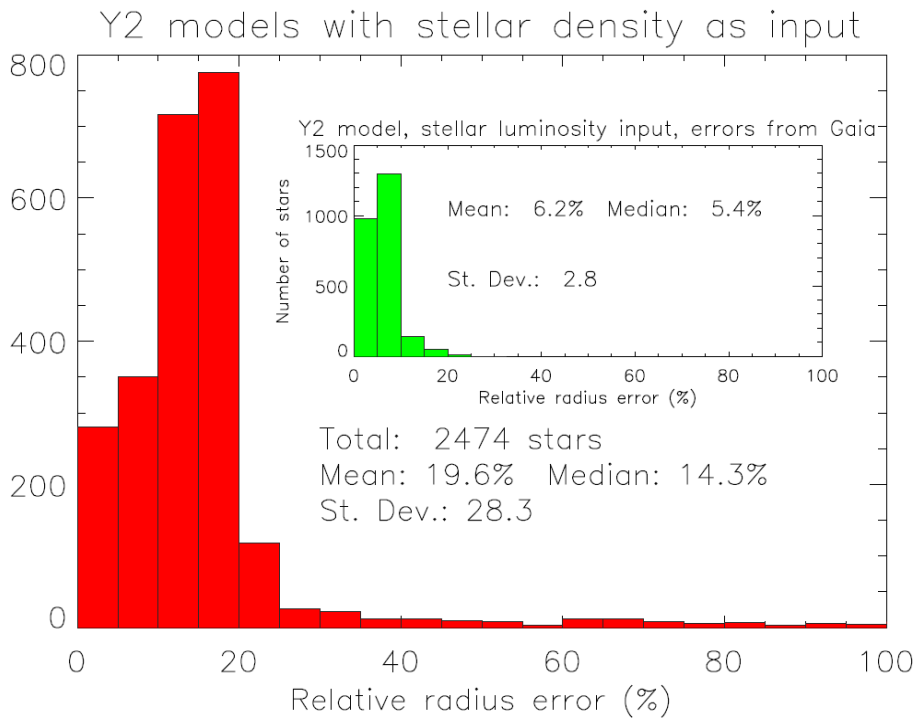}
\caption{Error distribution in stellar radius for a representative sample of Kepler transiting planet candidates derived from stellar evolution models 
using as input $T_\mathrm{eff}$ and $\varrho$ estimates from Kepler lightcurves (red histogram) or $T_\mathrm{eff}$ and $L$ estimates based on Gaia 
parallax determinations (green histogram). }
\label{fig:2}       
\end{figure}

Another way of estimating the impact of Gaia is to gauge the differences in stellar radius errors for transiting planet hosts as determined from stellar evolution 
models when they are fed with a given $T_\mathrm{eff}$ and either an estimate of the stellar density $\varrho$ from the transit lightcurve
\footnote{A proxy for $L$ routinely used in transiting systems parameters estimation when no direct distance estimate is available. See e.g. Sozzetti et al. (2007).} or 
an actual absolute luminosity estimate based on Gaia parallax determination. As an illustrative example, we show in Figure~\ref{fig:2} 
the comparison between the error in stellar radius for a sample of Kepler transiting planet candidates based on the Yonsei-Yale models 
(Yi et al. 2003; Spada et al. 2013) using as input $T_\mathrm{eff}$ (from the Kepler input catalog, KIC) and $\varrho$ from the lightcurve fits, or instead 
$T_\mathrm{eff}$ from KIC and an $L$ estimate based on Gaia parallax information (and expected precision). As one can see, a statistical 
improvements of a factor 3 on average would be possible. Such reduction in expected uncertainties on this fundamental stellar parameter 
would directly propagate in the planetary radius estimates. 

In summary, the wealth of Gaia data holds promise to significantly reduce (if not outrightly eliminate) one major source of 
potential confusion in atmospheric characterization measurements of transiting exoplanets carried out by EChO. The both accurate 
{\it and} precise knowledge of stellar radii will translate in more accurate and precise planetary radii. As a result, the more precisely 
determined planetary densities will allow for significantly reduced uncertainties in the inferred compositions. In turn, these 
results will crucially inform any atmospheric characterization measurements with EChO, particularly when it comes to super-Earths, 
for which, degeneracies in the models of their physical structure indicate a wide range of possible compositions for similar masses 
and radii (Seager \& Deming 2010, and references therein).

\section{Target Selection}

Gaia will be capable of measuring accurate astrometric orbits and masses for giant planets within approximately 0.3-4 AU of moderately bright 
($6<G<13$ mag, with the current baseline pushing the bright limit down to $G\simeq2$ mag) F-G-K stars out to $\approx200$ pc from the Sun (Casertano et al. 2008). 
The Gaia potential for detection and characterization of giant planets orbiting the reservoir of thousands of nearby low-mass M dwarfs within a $\sim30$ pc 
from the Sun is presented in detail in Sozzetti et al. (2014). The Gaia contribution to the EChO target list in terms of actual transiting 
(and non-transiting) exoplanets detected astrometrically is described in detail in Micela et al. (this volume). We focus here on three further elements 
of information provided by Gaia that can contribute to the optimal selection of EChO targets. 

1) ALL known transiting planet systems of potential interest for EChO will be probed by Gaia to find evidence for outer companions 
(massive giant planets, brown dwarfs, and very low-mass stars) in at least three ways. First, Gaia will be capable of detecting astrometrically 
wide-separation faint companions (giant planets and brown dwarfs) by measuring significant secular changes in the proper motion. Second, brighter, 
unresolved companions (low-mass stars) might be recognized as `variability induced movers' in Gaia astrometry. Third, Gaia might possibly directly 
(or partly) resolve existing wide-separation ($> 0.2$ arcsec) companions down to the $G=20$ mag survey limit. Combining the Gaia data with the 
available information from long-term radial-velocity monitoring and direct imaging, it will thus be possible to characterize the neighbourhood 
of each transiting planet host in order to identify any extremely red objects that, being quite bright at $5-15$ $\mu$m, might complicate the 
interpretation of EChO's atmospheric characterization measurements. 

2) It might be possible to detect hints of orbital motion due to the transiting planets themselves in Gaia astrometry, thus providing improved mass estimates 
through a combined radial-velocity + astrometric orbital solution, to further help in understanding EChO's atmospheric characterization measurements. 
The presently-known sample is characterized by objects that are too far away ($d\geq 100$ pc such as those 
in the Kepler field) and/or too close-in (few days of period, such as those detected in ground-based wide-field surveys) to produce astrometric signals above the 
best-case single-measurement performance by Gaia of $\sim20$ micro-arcsec ($\mu$as, see Sozzetti et al. 2014). One notable exception is HD 80606b, whose inferred 
astrometric signature ($\sim30$ $\mu$as) might be marginally detectable in Gaia data. The CHEOPS mission will monitor the sample of nearest stars 
($d\leq30$ pc) with planets detected by radial-velocity (RV) measurements and it might identify transiting planets with orbits of at least a few weeks of period 
that are massive enough to be producing detectable signals in Gaia astrometry, such as Gl 86b. The Gaia sampling pattern due to the scanning law will however 
complicate detections (in astrometry alone) for orbital periods significantly shorter than $\sim2$ months. This is the typical limiting value of orbital period for transiting 
systems that might be found by the TESS mission. The typical distance to stars in the TESS target sample ($d>50-100$ pc) will make any astrometric detection with Gaia 
very difficult, unless the transiting companion has a mass of at least a few tens of Jupiter masses, or larger. 

3) While the predefined Gaia scanning law will not allow to carry out intensive photometric monitoring of the potential EChO target sample, 
the mission will still deliver typical per-measurement precision of $\sim1$ milli-mag at the relatively bright magnitudes of interest for EChO. 
Particularly for targets observed more frequently by Gaia (with, e.g., $> 100$ measurements during the five-years nominal mission duration), 
it will thus be possible to combine Gaia photometric data with existing ground-based and space-borne (e.g., CHEOPS) photometry in order to characterize 
the various timescales of microvariability of the reservoir of potential EChO targets using a time baseline on the order of two decades. 





%

\begin{acknowledgements}
This work has been funded in part by ASI under contract to INAF I/022/12/0 (EChO Mission - Assessment Phase). 
An anonymous reviewer provided useful comments that materially improved the manuscript.
\end{acknowledgements}



\end{document}